\documentclass[a4paper,11pt]{article}
\pdfoutput=1 
\usepackage{jcappub}
\usepackage[utf8]{inputenc}
\usepackage{graphicx}
\usepackage{bm}
\usepackage{amsmath}
\usepackage{siunitx}
\usepackage{hhline}
\usepackage{multirow}
\usepackage{tabularx}
\usepackage{comment}
\usepackage[dvipsnames]{xcolor}
\usepackage{mathtools}
\usepackage[normalem]{ulem}
\usepackage{hhline}
\usepackage{gensymb}
\usepackage{makecell}

\newcommand{\dd}{\textrm{d}}
\definecolor{darkred}{rgb}{0.15, 0.15, 0.9}

\def\apj{\ref@jnl{ApJ}}                 

\title{A model-independent tripartite test of cosmic distance relations}

\author[*, a, b]{Isabela Matos,}
\author[*, b, c, d]{Miguel Quartin,}
\author[e]{Luca Amendola,}
\author[f]{Martin Kunz,}
\author[a,d]{Riccardo Sturani}

\affiliation[a]{Instituto de Física Teórica, Universidade Estadual Paulista \& ICTP South
American Institute for Fundamental Research, São Paulo 01140-070, SP, Brazil}

\affiliation[b]{Instituto de Física, Universidade Federal do Rio de Janeiro, 21941-972, Rio de Janeiro, RJ, Brazil}

\affiliation[c]{Observatório do Valongo, Universidade Federal do Rio de Janeiro, 20080-090, Rio de Janeiro, RJ, Brazil}

\affiliation[d]{PPGCosmo, Universidade Federal do Espírito Santo, 29075-910, Vitória, ES, Brazil}

\affiliation[e]{Institute of Theoretical Physics, Heidelberg University, Philosophenweg 16, 69120 Heidelberg, Germany}
	
\affiliation[f]{Département de Physique Théorique and Center for Astroparticle Physics,
Université de Genève, Quai E. Ansermet 24, CH-1211 Genève 4, Switzerland}

\affiliation[*]{\it These authors contributed equally to this work.}

\emailAdd{isabela.matos@ictp-saifr.org, mquartin@if.ufrj.br}

\abstract{
Cosmological distances are fundamental observables in cosmology. The luminosity ($D_L$), angular diameter ($D_A$) and gravitational wave ($D_{\rm GW}$) distances are all trivially related in General Relativity assuming no significant absorption of photons in the extragalactic medium, also known as cosmic opacity. Supernovae have long been the main cosmological standard candle, but bright standard sirens are now a proven alternative, with the advantage of not requiring calibration with other astrophysical sources. Moreover, they can also measure deviations from modified gravity through discrepancies between $D_L$ and $D_{\rm GW}$. However, both gravitational and cosmological parameters are degenerate in the Hubble diagram, making it hard to properly detect beyond standard model physics. Finally, recently a model-independent method named FreePower was proposed to infer angular diameter distances from large-scale structure which is independent of the knowledge of both early universe and dark energy physics. In this paper we propose a tripartite test of the ratios of these three distances with minimal amount of assumptions regarding cosmology, the early universe, cosmic opacity and modified gravity. We proceed to forecast this test with a combination of LSST and Roman supernovae, Einstein Telescope bright sirens and a joint DESI-like + Euclid-like galaxy survey. We find that even in this very model-independent approach we will be able to detect, in each of many redshift bins, percent-level deviations in these ratios of distances, allowing for very precise consistency checks of $\Lambda$CDM and standard physics. It can also result in sub-percent measurements of $H_0$.
}

\begin{document}

\maketitle

\section{Introduction}

The successful direct detection of gravitational waves (GWs) has delivered a revolutionary new tool to astronomy. GW standard sirens provide direct and self-contained distance measurements which skip all the usual steps in the cosmic distance ladder. These absolute distance indicators rise to the level of  standard candles when their electromagnetic counterparts are also detected because these provide a measurement of their redshift, a quantity otherwise broadly degenerate with source astrophysical parameters. For both the current and next generation ground-based GW detectors, these counterparts are believed to mostly be in the form of short gamma-ray burst (GRB) or kilonova (KN) explosions resulting from mergers of binary neutron star (BNS) systems and possibly also mergers of a neutron star and a black-hole. These multimessenger events are called bright standard sirens, in contrast with dark standard sirens, which rely on cross-correlation with galaxy surveys to constrain the redshift degree of freedom~\citep{Schutz:1986gp,DelPozzo:2011vcw,Chen:2017rfc,Bera:2020jhx,Finke:2021aom,LIGOScientific:2021aug,Mukherjee:2022afz}.

A number of other uses of GWs for cosmology without electromagnetic counterpart have been proposed, such as: through the assumption of understood features in the mass distribution of binary black holes (BBH) such as a characteristic mass scale or a mass-gap~\citep{Chernoff:1993th,Taylor:2011fs,Farr:2019twy,Leyde:2022orh,Ezquiaga:2022zkx}; through measurements of tidal deformation in the waveform~\citep{Messenger:2011gi}; or through priors on the redshift distribution of sources~\citep{Leandro:2021qlc}. In another approach, the use of bright siren clustering as probes of the density and peculiar velocity fields have also been  recently shown to be a very promising avenue~\cite{Palmese:2020kxn,Diaz:2021pem,Alfradique:2022tox}. We will nevertheless not consider these possibilities in this work.

Assuming General Relativity, bright standard sirens thus provide a direct reconstruction of the Hubble diagram. In particular, from its low redshift limit, it allows a measurement of the Hubble constant $H_0$~\cite{Schutz:1986gp} with minimal dependence on cosmology. The single bright siren detected by the first three runs of the LIGO/Virgo/KAGRA collaboration \cite{TheLIGOScientific:2014jea,TheVirgo:2014hva}, the gravitational wave GW170817~\citep{LIGOScientific:2017vwq} together with the GRB170817A, already provided a wealth of data, in particular a first GW measurement of $H_0$~\citep{LIGOScientific:2017adf} and a very precise measurement of the speed of propagation of GWs \cite{LIGOScientific:2018dkp}, which had great impact on extensions of the $\Lambda$CDM model from modified gravity theories and/or dark energy models \cite{Baker:2017hug, Creminelli:2017sry, Ezquiaga2017, Sakstein2017} (see, however, \cite{deRham:2018red}). With third generation GW detectors, such as the Einstein Telescope (ET)\footnote{\url{https://www.et-gw.eu/}} and the Cosmic Explorer (CE)\footnote{\url{https://cosmicexplorer.org/}}, the amount of multi-messenger events is expected to grow by orders of magnitude, potentially reaching the thousands in a few years, depending on the still poorly known rate of BNS coalescences, the number of concurrently operating GW observatories and the dedicated follow-up telescope infrastructure. In any case, tight constraints on beyond standard model scenarios are expected with these multimessenger events.

Modified gravity is a indeed a very suitable research case for these events since it can change the propagation of GW not only through the GW speed, but also modifying the inference of the GW luminosity distances $D_{\rm GW}$. In other words, although in General Relativity $D_{\rm GW}$ equates to the traditional electromagnetic luminosity distance $D_L$, in modified gravity they can differ at the same redshifts~\cite{Lombriser2016, Amendola2018, Belgacem:2018lbp, Nishizawa2018, Lobato:2022puv}. This can happen even if graviton number is conserved \cite{Lobato:2022puv}. Tests of the redshift dependence of the GW distance have been forecasted in many works for various phenomenological dark energy models, often assuming parametrizations for this deviation  (see, e.g.~\cite{Zhao2011, DAgostino2019, Nunes2019, Zhang2019, Nishizawa2019, Bachega2020,  Matos:2021qne, Allahyari2022}). However, to confidently detect any presence of modified gravity one would have to overcome the problem of the strong degeneracies between the background cosmological parameters ($H_0$, $\Omega_{m0}$, $w$, etc.) and the parameters controlling the deviation of $D_{\rm GW}/D_L$ from unity, as discussed in \cite{Matos:2022uew}. This implies that one cannot directly measure $H_0$ from standard sirens alone if one allows a completely general gravitational theory.

Leaving aside GWs, an analogous distance comparison tests Etherington's reciprocity-relation~\cite{Etherington1933}, which relates luminosity $D_L$ and angular diameter $D_A$ distance measurements. In metric theories of gravity, if the photon number is conserved, they are related through $D_L = D_A (1+z)^2$. A breakdown of this relation would entail the violation of photon number conservation or the Riemannian spacetime description of gravity. Disregarding the catastrophic latter option, a violation would imply that the Universe is not transparent on large scales, thereby inducing a change to the inferred $D_L$. Since GWs would be unaffected by this kind of phenomenon, this effect would also change the ratio $D_{\rm GW}/D_L$.

Tests of this duality relation have been carried out in the literature, typically by comparing luminosity distances of supernovae (SNe) and angular diameter distance measurements \cite{Bassett:2003vu,Avgoustidis:2009ai,Avgoustidis:2010ju,Yang2013,Liao:2015uzb, Hogg:2020ktc} or of different probes \cite{Holanda:2010vb,Holanda:2011hh,Holanda:2012at,Qi2019} to test either parametrized deviations from the relation or specific models. Refs.~\cite{Qi2019,Belgacem:2019zzu,Hogg:2020ktc} are especially interesting for us since they used GW standard sirens. Particularly \cite{Hogg:2020ktc} considered both SNe and GW sirens to measure the luminosity distance, and they also used a machine-learning based approach in addition to a parametrized model. 

In this paper we propose a model-independent tripartite test of distance relations, and investigate with which precision future observations could detect deviations among the distance relations. The tripartite test relies on ratios of $D_A,\,D_L,$ and $D_{\rm GW}$, obtained using large-scale structure, type Ia supernovae (SNe) and bright sirens, respectively, with minimal assumptions. In particular, to start we remark that this can be achieved without reliance on any parametrizations or considerations about the redshift dependence of these distances. Second, we show that one does not need calibrate the SNe distances with additional datasets. And finally, by relying on the recently proposed large-scale structure (LSS) FreePower method~\cite{Amendola:2019lvy,Amendola:2022vte,Amendola:2023awr,Schirra:2024rjq}, one can obtain measurements of $D_A$ which are independent of a model for both early universe physics and dark energy. 

To perform forecasts, for SNe we will rely on both the Roman Space Telescope\footnote{\url{https://www.stsci.edu/roman/observatory}} \cite{Spergel:2015sza},  which will be able to observe them up to $z \simeq 3$, and on the Rubin Observatory's Legacy Survey of Space and Time (LSST)\footnote{\url{https://www.lsst.org/}} \cite{LSSTScience:2009jmu}, which will see $\mathcal{O}(10^6)$ explosions up to $z \simeq 0.6$ with photometric redshifts. For LSS we will focus on forecasts for an Euclid-DESI-like spectroscopic galaxy survey, which is expected to yield comparable galaxy survey data to Euclid~\cite{laureijs2011euclid} and DESI~\cite{DESI:2016fyo}. In particular, for $z \le 0.6$ we will assume a DESI-like Bright Galaxy Survey (BGS) covering an area of 14000 deg$^{2}$ based on~\cite{Hahn:2022dnf}, while for $0.6 \le z \le 2.0$ we will assume an Euclid-like survey with an area of 15000 deg$^{2}$, based on~\cite{Amendola:2023awr}. Finally, for BNS GWs we will use the Einstein Telescope (ET).  ET will detect GWs from various types of sources up to very high redshifts, possibly $z \simeq 20$ \cite{Maggiore:2019uih, Sathyaprakash:2009xt}. In particular, BNS mergers are expected to be detectable up to $z \simeq 2$. The major difficulty will be the identification of electromagnetic counterparts, in view of the limited capability in sky localization of GW detectors alone, which might imply that bright standard sirens will be detected only in the range $z < 1$.

\section{Theory}

Our goal is to investigate to which precision one can measure independent ratios between the three distances $D_A,\,D_L,$ and $D_{\rm GW}$. Any two ratios define them completely, and we choose to work with these two:
\begin{equation}
    \zeta := \frac{D_{\rm GW}}{D_A(1+z)^2}\,, \quad \eta := \frac{D_{L}}{D_A(1+z)^2}\,.  \label{xi_eta}
\end{equation}
This choice is interesting because the presence of cosmic opacity or violation of the reciprocity relation is encoded in $\eta$, while a modification of the propagation of GWs due to e.g. modified gravity, is captured by $\zeta$. Both are equal to unity in the standard model, with minimal assumptions.  The third ratio, between GW and luminosity distances is often denoted by the symbol $\Xi$ in the literature, such that $\Xi = \zeta/\eta$. Ignoring cosmic opacity, one recovers $\Xi=\zeta$. However, since we want to be as general as possible, we chose to split the two contributions and use different symbols to avoid confusion.  In particular, as discussed in the introduction, we will avoid having to specify a cosmological model or parametrizations for $\zeta$ and $\eta$.

\subsection{Cosmic opacity}\label{subsec:opacity}

Regardless of the assumed theory of gravity, in any universe where light propagates along null geodesics and photon number is conserved, the Etherington distance duality relation holds. That is, the luminosity distance relates to the angular diameter distance so that the cosmic opacity parameter $\eta$ is always unity. If, on the other hand, photons are lost on the way from the source to the observer, this relation is violated, since the inferred luminosity distance will change. This can happen, for instance, due to  intergalactic dust or in models where photons decay into axions in the presence of magnetic fields \cite{Kunz:2004ry, More2009cosmic, Jaeckel:2010ni, Avgoustidis:2010ju, Tiwari2017}. In this case  the GW propagation is not affected and a comparison of GW and luminosity distances alone would indicate a deviation that we can not distinguish from modified gravity since these two effects can have similar redshift dependence.

The angular diameter distance, on the other hand, is a pure geometric quantity that is not affected by cosmic opacity or by the purely gravitational degrees of freedom (assuming a given background cosmology). As we will discuss below, it can be obtained from galaxy surveys (up to a $H_0$ factor) without assuming a model or knowledge of the early universe physics. Thus, a comparison of $D_A$ with $D_L$ at various redshift bins can constrain deviations of the relation $\eta = 1$ in a model-independent way.

\subsection{Modified GW propagation}
\label{subsec:Modified_GW}

Although we will not be restricted to this case, we will be particularly interested in constraining modified gravity (MG) scenarios that affect the GW distance, and use this test to measure the effective Planck mass and its running in this context. More precisely, the MG scenario we will investigate is the large class of scalar-tensor theories described in~\cite{Bellini2014}, which corresponds to the set of effective field theories of dark energy. Our cosmology is defined by a perturbed spatially flat FLRW metric. In the Newtonian gauge, we can write
\begin{equation}
    ds^2 = a(\tau)^2\left[ -(1 + 2\Psi)d\tau^2 + (1 - 2\Phi)(\delta_{ij} + h_{ij})dx^idx^j\right]\,,
\end{equation}
where $a$ is the scale factor, $\tau$ is the conformal time, $\Phi$ and $\Psi$ are the scalar potentials and $h_{ij}$ is the transverse traceless tensor perturbation, that can be decomposed into $h_+$ and $h_{\times}$ polarizations. In short, as argued in \cite{Bellini2014, Gleyzes:2014qga, Gubitosi:2012hu}, the evolution of all linear perturbations is completely determined once we specify the background solution $\mathcal{H}(\tau)$ and four time-dependent functions 
\begin{equation}
    \left(\alpha_M, \alpha_T, \alpha_K, \alpha_B\right)\,.
\end{equation}
The propagation of the tensor modes is, however, only sensitive to the first two of these functions, since
\begin{equation}
    h_P'' + (2 + \alpha_M){\mathcal H}h'_P + (1 + \alpha_T)k^2h_P = \Pi_P\,, \label{GW}
\end{equation}
with $P = +, \times$, where $\Pi_P$ is the tensor part of the anisotropic stress of matter.

The very precise measurement of the speed at which gravitational waves (GWs) propagate implies $|\alpha_T| < 10^{-15}$ at low redshifts~\cite{Creminelli:2017sry,Baker:2017hug,Sakstein2017, Ezquiaga2017}, and so we set it to zero. Then one can see that the only MG signature in the GW propagation is the modified friction $\alpha_M$, which is defined as the running of an effective Planck mass in these theories,
\begin{equation}
    \alpha_M := \frac{d(\ln M_{\ast}^2)}{d\ln a}\,.
\end{equation}
Such modified damping rate changes the value of the GW amplitude detected by local interferometers. Since $h_P$ falls with the distance to the source, one can absorb the modified friction $\alpha_M$ into the distance and define a \textit{GW distance}, which relates to the luminosity distance via
\begin{equation}
    \frac{D_{\rm GW}(z)}{D_L(z)} = \exp\left\{\frac{1}{2}\int_0^z \frac{\alpha_M(\tilde{z})}{1 + \tilde{z}}d\tilde{z}\right\} = \sqrt{\frac{M_{\ast}^2(0)}{M_{\ast}^2(z)}}\,. \label{gw_distance}
\end{equation}

The modified friction in the GW propagation that leads to a new cosmological distance is a feature present not only in the context of \cite{Bellini2014}, but also in other classes of modified gravity models. Also, we notice that GW and luminosity distances differ even if gravitons are conserved along null geodesics in various models. This is due to the fact that such distance is inferred from the measured amplitude of the GW, and not from its energy density, unlike photons, as discussed in \cite{Lobato:2022puv}. This is why photon and graviton conservation have different impacts on the respective distance estimates. Finally, we also note that the MG models considered here change not only the GW propagation, but also growth of structure and gravitational lensing \cite{Saltas:2014dha}. A comparison of how future observations constrain these modifications can for example be found in \cite{Matos:2022uew}. In this paper, however, we are interested in fully model-independent constraints from three different ways to measure distances, which is why we do not consider the impact of MG models on the growth of the large-scale structure.

We can see from Eq.~\eqref{gw_distance} that in the context of models with a time-varying Planck mass, the GW distance is changed by the ratio of the Planck mass at emission and observation. Many models with evolving Planck mass also implement screening to avoid being ruled out by solar system constraints on modifications of General Relativity, and some of the screening mechanisms would ``hide'' the global value of the Planck mass, and thus suppress the change in GW distance. However, this is strongly scenario dependent, as there are screening mechanisms that do not affect the Planck mass (e.g.~Vainshtein screening~\cite{Kimura:2011dc}) and also models that do not require screening (e.g.~the non-local models of \cite{Belgacem:2018lbp}). Here we want to discuss more generally the constraints that we can obtain on relative distance measurements to {\em test} whether the distances are the same, and so we do not enter into a detailed discussion of these different models.

Although tests of the GW-distance-redshift relation have been proposed in literature using multimessenger events, there is a degeneracy between  the redshift evolution of the gravitational and background cosmology parameters. Since our aim is to have a model-independent test, we instead propose measuring $\zeta$ by directly comparing the two cosmological distances $D_{\rm GW}$, $D_L$ (up to a constant) at several redshift bins, using standard sirens and (non-calibrated) supernovae. When allowing for independent deviations in the luminosity distance, we instead propose comparing $D_{\rm GW}$ to $D_A(1+z)^2$ since this isolates the modified gravity effects from a possible cosmic opacity.

\section{Data}

\begin{figure}
    \centering
    \includegraphics[width=0.48\linewidth]{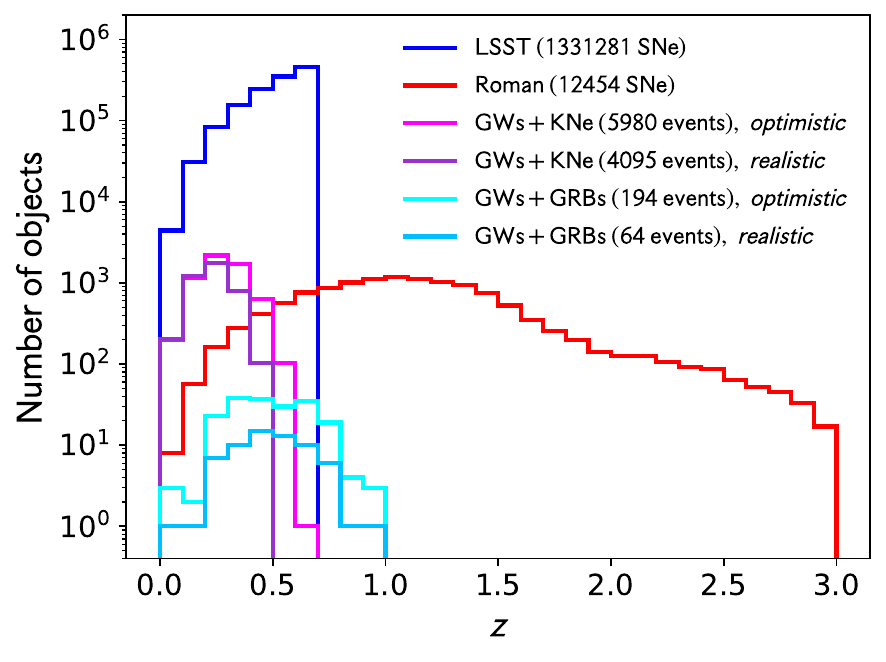}
    \includegraphics[width=0.49\linewidth]{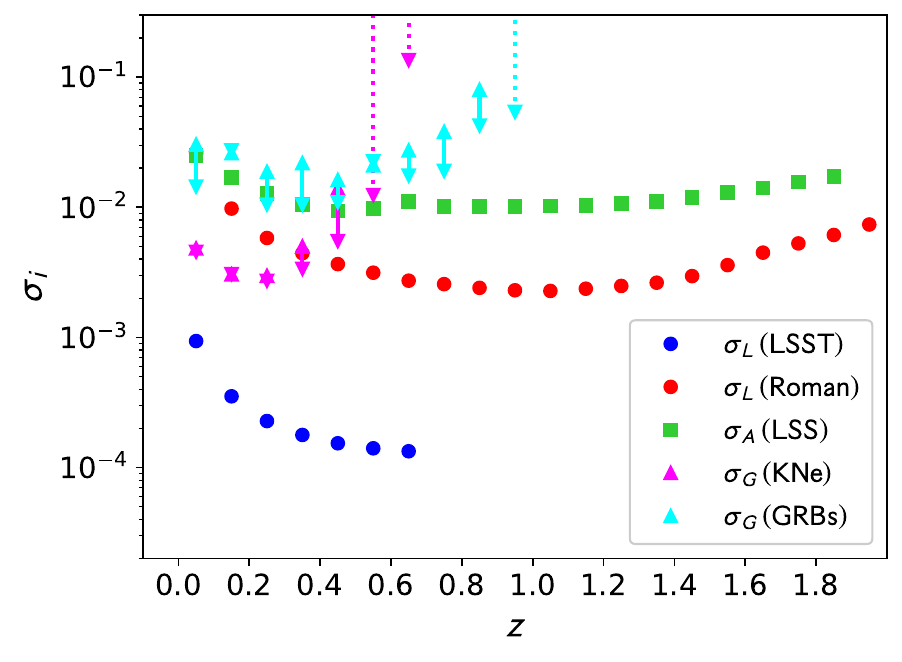}
    \caption{\emph{Left:} number of objects in each of the forecasts considered as a function of redshift [see text]. \emph{Right:} relative errors in the distances. Arrows connect realistic to optimistic scenarios.
    }
    \label{fig:data}
\end{figure}

\subsection{Supernovae}

There are two main upcoming SNe surveys: LSST, which will detect hundreds of thousands of events at low and intermediate redshifts, and Roman that should detect SNe all the way to $z=3$. We assume supernova distances will be measured with magnitude uncertainties given by the sum in quadrature of the intrinsic scatter $\sigma_{\rm int} = 0.13$ mag with both the lensing-induced scatter of $\sigma_{\rm lens} = 0.052 z$~\cite{Quartin:2013moa} and the peculiar velocity scatter $\sigma_{\rm pv} = (5/\log 10)\sigma_v/(c z)$, for $\sigma_v = 250\,$km/s. For the number of events for LSST, since ET will start a decade later we assume a 10-year SN survey over an 18000 deg${}^2$ area with a constant 15\% SN completeness in the redshift range $0 < z < 0.7$.\footnote{This yields the same SN catalog as a 5-year survey with 30\% completeness, as considered in~\cite{Quartin:2021dmr}. A more careful estimate of the completeness was performed in~\cite{Garcia:2019ita} using a public LSST survey strategy under consideration, which showed that a 15\% completeness was achievable for $z\le 0.5$.} We assume spectroscopic follow-up of all these events.  For Roman, we follow~\cite{Rose:2021nzt} assuming the combination of both WIDE and DEEP surveys.

\subsection{Bright standard sirens}

We consider two different forecasts of bright standard sirens that are shortly described below. Both of them corresponds to an effective 5-year  Einstein Telescope observation run (with 80\% duty cycle) and differ in the type of electromagnetic counterpart expected to be seen with different instruments.
\begin{itemize}
    \item [(1)] GWs+KNe: we take the forecasts of \cite{Alfradique:2022tox} of joint detections of GWs with the Einstein Telescope and kilonovae (KNe) with the Vera Rubin observatory.
    \smallskip
    \item [(2)] GWs+GRBs: we forecast joint detections of GWs with the Einstein Telescope together with short gamma-ray bursts (GRBs) emitted by binary neutron stars (BNS).
\end{itemize}

An intrinsic limitation in GW distance estimates is the degeneracy in the waveform between the effects of distance and inclination, the latter being the angle between the binary's orbital angular momentum and the line of sight. This degeneracy is partially broken when a short GRB follow-up is detected, a signal strongly affected by the orientation of the binary. For low inclination angles (i.e., a jet pointing at us), however, the analysis based on the Fisher matrix alone can be unreliable due to this degeneracy.

The distance error estimates can be obtained in at least two ways to deal with this issue. One is the analytical likelihoods obtained by~\cite{Chassande-Mottin:2019nnz} as an extension of~\cite{Cutler:1994ys} by assuming independence between extrinsic and intrinsic parameters in the waveform. These likelihoods were proven to be accurate even for low inclinations. However, this approximation breaks down for GWs close to the plane of the ET, for which both polarizations are degenerate. The second alternative is to go beyond the Fisher approximation using the Derivative Approximation for Likelihood (DALI) method developed in~\cite{Sellentin:2014zta}. Recently, a DALI code for GW, dubbed GWDALI,\footnote{\url{https://gwdali.readthedocs.io/en/latest/}} was implemented in \cite{Desouza2023}. Both DALI and the analytic approximation of \cite{Chassande-Mottin:2019nnz} should provide compatible results (see \cite{deSouza:2023gjv}), and we will use both below.

The mock data of BNS mergers with KNe (1) is described in detail in \cite{Alfradique:2022tox}, and thus here we only summarize the most relevant aspects for our purposes. For this data set we used the analytic likelihood of \cite{Chassande-Mottin:2019nnz}. Forecasting the number of events highly depends on the assumption of a merger rate per comoving volume in the source frame. We assume a local rate of $\mathcal{R}_0 = 300 \; \mathrm{Gpc}^{-3} \mathrm{yr}^{-1}$ with a redshift-dependence proportional to the Madau-Dickinson star formation rate~\cite{Madau:2014bja} (neglecting the time-delay between star formation and binary merger). This is compatible with the loose LIGO GWTC-3 bounds~\cite{LIGOScientific:2021psn}.

The SNR threshold for GW detection with the three interferometers of the Einstein Telescope has been set to 12. The follow-up was assumed to be performed by the Rubin observatory using an optimized strategy which relies on the distance estimates to decide the integration time. Two scenarios were considered:  
\begin{itemize}
    \bigskip
    \item Realistic: 10\% of the total usable telescope time is dedicated to GW follow-up;
    \item Optimistic: 50\% of the total usable telescope time is dedicated to GW follow-up.
\end{itemize}

For the GRB forecasts (2), due to the low inclination angles of the binaries, we do not rely on Fisher Matrix codes or on the likelihood of \cite{Chassande-Mottin:2019nnz}, and neither do we use the inverse SNR as an estimate of the error in the distance (like~\cite{Belgacem2019}). Instead, we use GWDALI. We choose to work with only the first correction to the Fisher matrix  -- the \textit{doublet} -- to avoid numerical instabilities that might occur in higher terms due to the presence of higher derivatives of the waveform. 

We also simulate a population of GW events emitted by coalescing BNS following~\cite{Alfradique:2022tox} as for BNS. We then proceed as in~\cite{Belgacem2019} regarding the detectability of the follow-up. We assume a future THESEUS-like telescope \cite{THESEUS:2017wvz}, with flux limit of $0.2 \; \text{ph sec}^{-1}\text{cm}^{-2}$ in the 50–300 keV band, and the Gaussian jet profile of GRB170817A. Imposing this flux limit, we further selected among the GW events those with a detectable GRB. The counterpart selection strongly restricts the observable inclination angles $\iota$ to be lower than around $23\degree$. We can thus use this information as a prior on $\iota$ when estimating the errors on the several parameters and partially break its degeneracy with the GW distance, improving the error measurements on the latter as compared to the KNe case.

We also consider two scenarios for joint GW and GRB detections according to Table 2 of~\cite{Belgacem2019}:
    \begin{itemize}
        \item Realistic: only includes events whose sky localization error is lower than 5$\degree$, for which redshift measurements are more likely to happen (1/3 of all events);
        \item Optimistic: all events whose GRB will be detected are assumed to have their redshifts determined.
    \end{itemize}

\subsection{Galaxy power spectrum and bispectrum}

The FreePower method was originally proposed in~\cite{Amendola:2019lvy,Amendola:2022vte} but only named in~\cite{Amendola:2023awr}. More recently, it was applied to BOSS data in~\cite{Schirra:2024rjq}. It allows an estimate of some cosmological functions 
with no need of modelling the power spectrum shape and background evolution, thus regardless of knowledge of both the early universe physics and dark energy models. In particular, in~\cite{Amendola:2023awr} the one-loop power spectrum ($P$) and the tree-level bispectrum ($B$) were employed to derive model-independent forecasts for an Euclid-like survey on parameters like the growth rate $f(z)$, the dimensionless expansion rate $E(z)=H(z)/H_0$ and, what concerns us here, on the dimensionless angular-diameter distance 
\begin{equation}
    L_A(z) := H_0 D_A(z) \,.    
\end{equation}
Here we extend our galaxy survey to include a low-redshift DESI BGS galaxies, i.e., we forecast for a joint DESI-like + Euclid-like survey.

The power spectrum and bispectrum depend on the linear power spectrum, on the linear growth function, and on two bias functions and five so-called bootstrap parameters~\cite{D'Amico_2021}. Bias and bootstrap parameters depend only on redshift. All these quantities have been left free to vary in a very general way. The bootstrap parameters characterize the higher-order kernels and have been obtained by imposing general properties, namely  the equivalence principle and  the conservation of the matter energy-momentum tensor. The constraints on $E(z), L_A(z)$ come from the Alcock-Paczyński effect on the combination of the power spectrum monopole with higher multipoles,\footnote{or, likewise, on the combination of different $\mu$ bins, also called wedges.} which in the absence of AP are simply due to the redshift-space distortions (RSD). They therefore depend only on statistical isotropy~\cite{Amendola:2023awr}. In particular, this method does not require any calibration with the inverse distance ladder, differing from  measurements of angular diameter distances from BAO, after fixing the sound horizon scale. In the FreePower method, no knowledge of the early Universe or the location of the BAO wiggles is required. In fact, the method has been shown to work even if there were no wiggles.  However, we remark that the resulting constraining power does depend on the actual power spectrum shape (which in a forecast means it depends on the fiducial cosmology), even if that does not need to be known a priori. Finally, we note that on linear scales there is a degeneracy between $E(z), L_A(z)$, which is only broken by considering non-linearities, here modelled with the one-loop terms.

For our forecasts, we make use of the one-loop power spectrum and the tree-level bispectrum. Since FreePower forecasts depend strongly on the choice of the cut-off scale $k_{\rm max}$, care must be taken for this choice. For both DESI-like ($z \le 0.6$) and Euclid-like forecasts ($z \ge 0.6$), we use $k^P_{\rm max}=0.25 \,h/$Mpc for $P$ and $k^B_{\rm max}=0.1 \,h/$Mpc for $B$. Concerning the fiducial values of the cosmological and bias parameters, for the Euclid-like survey we follow in general~\cite{Amendola:2023awr}, while for the DESI-like BGS survey, we follow a similar approach except that for the linear bias we assume $b_{1,\rm BGS}(z) = 1.34 / D_+(z)$, where $D_+(z)$ is the linear growth function. For DESI we also use the number density estimates provided in~\cite{Hahn:2022dnf}.

In Appendix~\ref{sec:freepower} we provide more details on the FreePower method and on the choices made in its implementation for  this work (see also~\cite{Amendola:2022vte,Amendola:2023awr}).

\section{Methods}

Supernova data alone, without absolute distance calibration, can only measure the luminosity distance up to a constant factor which depends on the absolute magnitude $M_B$ of the supernova, and is degenerate with the Hubble constant $H_0$. To wit, supernova data measures $\lambda D_L$, where $\lambda$ is given by (using units of Mpc):
\begin{equation}
    \lambda \, := \, 10^{5+M_B/5}\,.
\end{equation}
This means we can only determine cosmic opacity up to a constant. 

With $N$ redshift bins, we would have, in principle, $3N$ data points, the values of $D_{\rm GW}$, $ \lambda D_L$ and $L_A$ at each bin. However, we do not want to specify the cosmological background, implying we have to rely only on the $2N$ measurable independent distance ratios $\tilde\eta_i$ and $\tilde\zeta_i$: 
\begin{equation}
    \begin{aligned}
    & \zeta_i = H_0 \frac{D_{GWi}}{L_{Ai}(1 + z_i)^2} =: H_0 \, \tilde\zeta_i  \,, \\ 
    & \eta_i = \frac{H_0}{\lambda} \frac{\lambda D_{Li}}{L_{Ai}(1 + z_i)^2} =: \frac{H_0}{\lambda}  \,\tilde\eta_i\,.
    \end{aligned}
\end{equation}
With the uncalibrated supernovae we have $2N + 2$ parameters to measure: $\zeta_i, \, \eta_i, (i=1,\dots,N)$, $\lambda$ and $H_0$. Therefore it is necessary to make extra assumptions in order to transform the covariance of the distances into the covariance of the parameters we want to forecast. 

It is natural to expect that cosmic opacity should be an integrated effect that accumulates over a range of redshifts, even without restricting the analysis to a specific model for how photons are lost along the line of sight. Thus, $\eta_1 = 1$ is a reasonable assumption, and we will adopt it in all cases. One can, then, constrain $\lambda/H_0$ using the first redshift bin: $\lambda/H_0 = \tilde{\eta}_1$. Of course one can also assume the simpler scenario in which there can be no possible cosmic opacity at all, and thus one can use the ratios $\tilde{\eta}_i$ in all redshift bins to calibrate the SNe and get $\lambda/H_0$. 

We now need one more constraint to close the system, but this choice is less obvious. From Eq.~\eqref{gw_distance}, we notice that an equality at redshift $z\rightarrow 0$ also holds for GW and angular diameter distances in the MG framework of section \ref{subsec:Modified_GW}. However, we do not know how fast these two distances deviate from each other. So one possible assumption is that MG is causing larger deviations at $z > \Delta z$ than at $z < \Delta z$, which allows the measurement of $H_0$ (and $M_B$), and thus, $\eta$ and $\zeta$ in all remaining bins. Likewise, the opposite assumption would be that MG is not relevant at high $z$, which entails $\zeta = 1$ for these redshifts, but this requires detecting high-$z$ standard sirens. Another option would be to measure $H_0$ independently locally with the traditional local distance ladder tools. Here we do not pursue this option due to the current tension in $H_0$. If one moreover assumes GR, even tighter constraints can be placed on both $H_0$ and $M_B$. In this work we thus study three possible alternative assumptions.

\begin{itemize}
    \item Case 1: $M_{\ast}^2$ varies slowly in the first redshift bin ($|\alpha_{M0}|\Delta z \ll 1$) and no photon was lost in the first bin. That is,
    \begin{equation}
        D_{\rm GW} = D_{L} = D_A \quad \text{for} \quad z < \Delta z \quad \Rightarrow \quad \zeta_1 = \eta_1 = 1\,.
    \end{equation}
    \item Case 2: $M_{\ast}^2$ varies slowly in the first redshift bin and there is no cosmic opacity in all bins. That is,
    \begin{equation}
        D_{\rm GW} = D_{L} \quad \text{for} \quad z < \Delta z \quad \text{and DDR is valid} \quad \Rightarrow \quad  \zeta_1, \eta_1, \cdots, \eta_N = 1\,.
    \end{equation}
    \item Case 3: Assume GR and no cosmic opacity, such that $\eta = \zeta = 1$ in all bins. With this we only measure $H_0$ and $\lambda$.
\end{itemize}

We can assume that the luminosity and GW distances in different redshift bins are uncorrelated. The angular diameter distances obtained with the FreePower method, however, have been shown to have non-negligible correlations between the different redshift bins~\cite{Amendola:2023awr}. While for an Euclid-like survey the average correlation was around 0.34 in~\cite{Amendola:2023awr}, here the addition of a DESI-like BGS survey at low $z$ breaks other degeneracies and the average correlation is lower, to wit 0.18. The covariance matrix of the ratios $(\tilde{\zeta}, \tilde{\eta})$ can be written as
\begin{align}
    &\text{Cov}(\tilde{\zeta}_i, \tilde{\zeta}_j) = \tilde{\zeta}_i \tilde{\zeta}_j C^A_{ij}\,, \qquad \text{Cov}(\tilde{\eta}_i, \tilde{\eta}_j) = \tilde{\eta}_i \tilde{\eta}_j C^A_{ij}\,, \qquad \text{for} \; i \neq j\,, \\
    &\text{Cov}(\tilde{\zeta}_i, \tilde{\eta}_j) = \tilde{\zeta}_i \tilde{\eta}_j C^A_{ij}\,, \\
    &\sigma_{\tilde{\zeta}_i}^2 = \sigma _{G,i}^2 + \sigma_{A,i}^2\,,\, \qquad \sigma_{\tilde{\eta}_i}^2 = \sigma _{L,i}^2 + \sigma_{A,i}^2\,,
\end{align}
where 
\begin{equation}
    C^A_{ij} = \text{Cov}(L_{Ai}, L_{Aj})/L_{Ai}L_{Aj} = \text{Cov}(\ln L_{Ai}, \ln L_{Aj})\,.
\end{equation}
Here $\sigma_{L,i}$, $\sigma_{A,i}$ and $\sigma_{G,i}$ are the relative errors for $\lambda D_{L}$, {\color{blue} $L_A$} and $D_{\rm GW}$ at $z_i$, respectively, and $\sigma_{\tilde{\zeta}_i}$ and $\sigma_{\tilde{\eta}_i}$ are the relative errors of the data points $\tilde{\zeta}_i$ and $\tilde{\eta}_i$.

We can then translate these into a Fisher matrix for the  2N+2 variables $\zeta$, $\eta$, $\lambda$ and $H_0$, whose components are:
\begin{align}
    &\mathcal{F}_{\zeta_i \zeta_j} = \frac{1}{H_0^2}[\text{Cov}^{-1}]_{\tilde{\zeta}_i \tilde{\zeta}_j}\,, \quad
    \mathcal{F}_{\eta_i \eta_j} = \frac{\lambda^2}{H_0^2} [\text{Cov}^{-1}]_{\tilde{\eta}_i \tilde{\eta}_j}\, \quad \mathcal{F}_{\lambda \lambda} = \sum_{k,l} \frac{\tilde{\eta}_k \tilde{\eta}_l}{\lambda^2H_0^2} [\text{Cov}^{-1}]_{\tilde{\eta}_k \tilde{\eta}_l}\,, \\
    &\mathcal{F}_{\zeta_i \eta_j} = \frac{\lambda}{H_0^2} [\text{Cov}^{-1}]_{\tilde{\zeta}_i \tilde{\eta}_j}\,, \quad 
    \mathcal{F}_{\zeta_i \lambda} = \sum_k \frac{\tilde{\eta}_k}{\lambda H_0^2} [\text{Cov}^{-1}]_{\tilde{\zeta}_i \tilde{\eta}_k}\,, \quad
    \mathcal{F}_{\eta_i \lambda} = \sum_k \frac{\tilde{\eta}_k}{H_0^2} [\text{Cov}^{-1}]_{\tilde{\eta}_i \tilde{\eta}_k}\,, \\
    &\mathcal{F}_{H_0 H_0} = \sum_{k,l} \left( \tilde{\zeta}_k\tilde{\zeta}_l [\text{Cov}^{-1}]_{\tilde{\zeta}_k \tilde{\zeta}_l} + \tilde{\eta}_k\tilde{\eta}_l [\text{Cov}^{-1}]_{\tilde{\eta}_k \tilde{\eta}_l} + 2\tilde{\zeta}_k\tilde{\eta}_l [\text{Cov}^{-1}]_{\tilde{\zeta}_k \tilde{\eta}_l}\right) \\
    &\mathcal{F}_{\zeta_i H_0} = -\frac{1}{H_0^2}\sum_k \left( \tilde{\zeta}_k [\text{Cov}^{-1}]_{\tilde{\zeta}_k \tilde{\zeta}_i} + \tilde{\eta}_k [\text{Cov}^{-1}]_{\tilde{\eta}_k \tilde{\zeta}_i} \right)\,, \\ & \mathcal{F}_{\eta_i H_0} = -\frac{\lambda}{H_0^2}\sum_k \left( \tilde{\zeta}_k [\text{Cov}^{-1}]_{\tilde{\zeta}_k \tilde{\eta}_i} + \tilde{\eta}_k [\text{Cov}^{-1}]_{\tilde{\eta}_k \tilde{\eta}_i} \right)\,,\\  &\mathcal{F}_{\lambda H_0} = -\frac{1}{\lambda H_0^3}\sum_{kl} \left( \tilde{\zeta}_k \tilde{\eta}_l [\text{Cov}^{-1}]_{\tilde{\zeta}_k \tilde{\eta}_l} + \frac{\tilde{\eta}_k \tilde{\eta}_l}{\lambda} [\text{Cov}^{-1}]_{\tilde{\eta}_k \tilde{\eta}_l}  \right)\,.
\end{align}
where $\lambda$ and $H_0$ here are evaluated at the best fit value.

We can also measure the running of the Planck mass, $\alpha_M$, if we assume it to vary slowly compared to the bin width $\Delta z$. We then have
\begin{equation}
    \alpha_{Mi} := \alpha_M(z_i + \Delta z/2) = \frac{2}{\Delta z}\left( 1 + z_i + \frac{\Delta z}{2}\right) \ln \left(\frac{\zeta_{i+1}}{\zeta_i}\right)\,,
\end{equation}
and thus, 
\begin{equation}
    \text{Cov}(\alpha_{Mi}, \alpha_{Mj}) = \frac{4}{(\Delta z)^2}\left( 1 + z_i + \frac{\Delta z}{2}\right)\left( 1 + z_j + \frac{\Delta z}{2}\right)\sum_{k = i}^{i+1}\sum_{l = j}^{j+1} \frac{\text{Cov}(\zeta_k, \zeta_l)}{\zeta_k \zeta_l}\,.
\end{equation}

\section{Results}

\subsection{Model-independent forecasts}

We now present the results of the forecasts in the model-independent scenario where both cosmic opacity and modified GW propagation are simultaneously allowed. The main result is that we will be able to test each of these two hypotheses with percent level precision even at the highest redshifts considered, the only assumption being that the three distances coincide at the lowest redshifts ($z < 0.1$).

\begin{figure}[t]
    \centering
    \includegraphics[width=\linewidth]{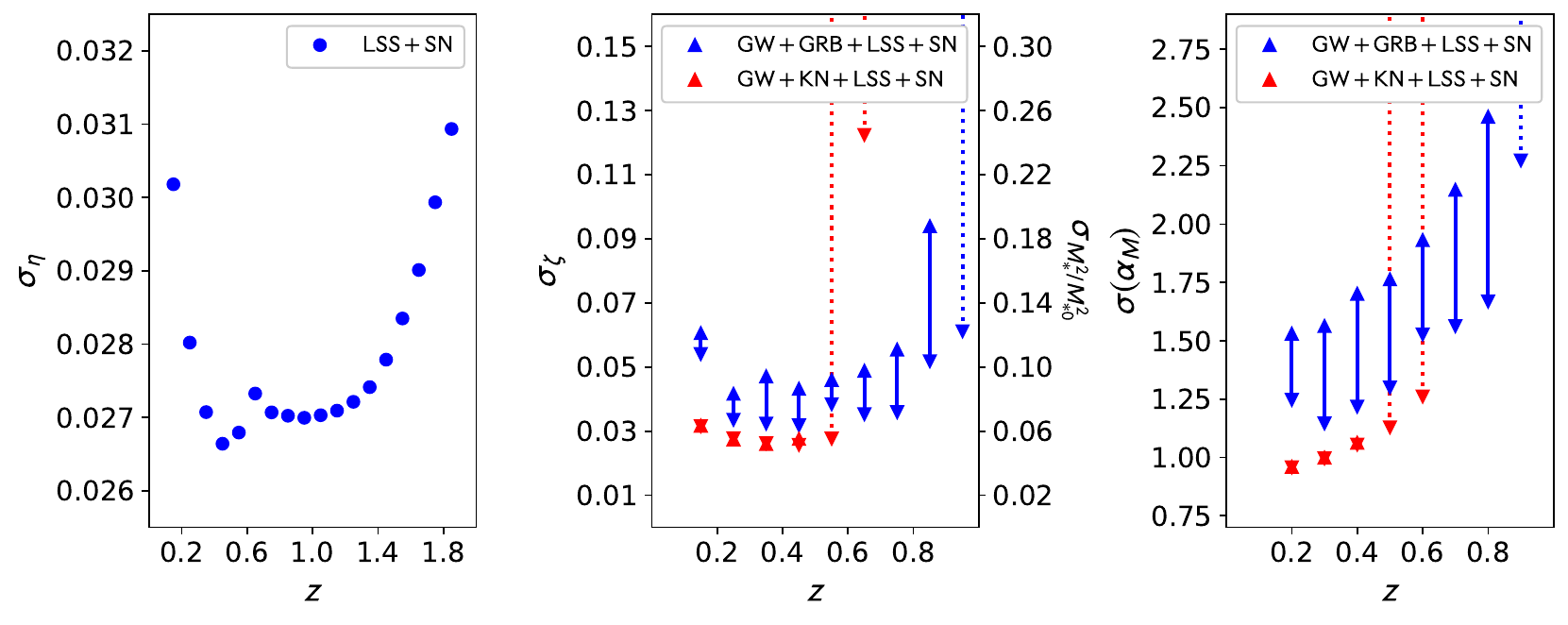}
    \caption{Forecasts uncertainties assuming $D_{A} = D_L = D_{\rm GW}$ at $z = 0$ (case 1). \emph{Left:} cosmic opacity parameter, which is independent of the bright siren data set.  \emph{Middle:} ratio $D_{\rm GW} / \big[D_A (1+z)^2$\big] and corresponding error in the Planck Mass normalized by its value today. \emph{Right:} time variation of the Planck mass $\alpha_M(z)$. Arrows connect optimistic to realistic scenarios. The $\eta$ error bars are highly correlated. 
    }
    \label{fig:error_xi_eta}
\end{figure}

The two first panels of Figure~\ref{fig:error_xi_eta} show the relative errors on the quantities $\eta$ and $\zeta$ in each redshift bin of size $\Delta z = 0.1$ in the general case (case 1). The former is obtained up to $z = 2$ while the latter up to $z = 1$ due to the lack of bright sirens beyond $z=1$. Assuming Eq.~(\ref{gw_distance}) holds, we also show the relative errors in the Planck mass ratio $M_{\ast}^2/M_{\ast0}^2$ in the second panel of Figure~\ref{fig:error_xi_eta}, and in the third panel, we show the errors for the derived quantity $\alpha_M$.

\begin{figure}[t]
    \centering
    \includegraphics[width=\linewidth]{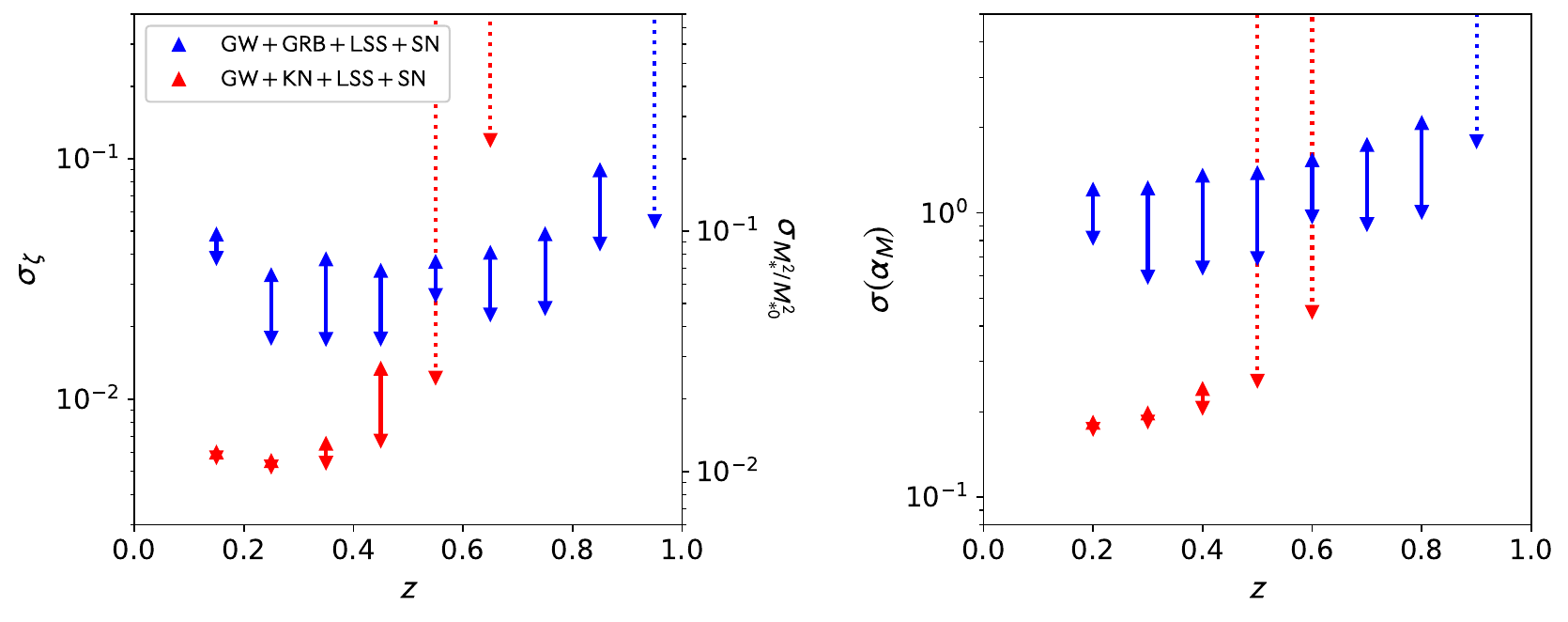}
    \caption{
    Similar to Figure~\ref{fig:error_xi_eta} assuming the reciprocity relation (no cosmic opacity). In this case $\Xi(z) = \zeta/\eta = \zeta = D_{\rm GW} / D_L$. Errors using KN improve substantially on this case.
    }
    \label{fig:error_xi_no_co}
\end{figure}

Figure~\ref{fig:error_xi_no_co} shows the forecasts for the relative error in $\zeta$ and the derived results for the Planck mass ratio and its running, assuming the reciprocity relation holds (case 2). In this case we obtain that any modification in the accumulated GW friction could be detected with less than 1\% precision from $z=0$ up to $z=0.4$ due to the high number of KNe observations, and with less than 10\% up to $z=0.9$ even in the realistic scenarios.

\begin{table}[t]
    \small
    \centering
    \setlength{\extrarowheight}{3pt}
    \begin{tabular}{l|c|c}
    \hhline{= = =}
      Data  & Constraint & Limitations\\  		\hline
      Lunar Laser Ranging \cite{Williams2004, Tsujikawa2019} &  $\alpha_{M0} < 0.02$ & local measurement \\
      Bright siren GW170817 \cite{Belgacem:2018lbp} & $\alpha_{M0} = 15.6^{+36.8}_{-19.4}$ & \makecell{local measurement} \\
      Planck \cite{planck2018} & \thead{$\alpha_{M0} = -0.040^{+0.041}_{-0.016}$ \\ $\beta = 0.72^{+0.38}_{-0.14}$} & \makecell{Assumes $\alpha_M = \alpha_{M0}a^{\beta}$; specific $\alpha_i$}\\
      \hline
    \end{tabular}
    \caption{Some of the current measurements of $\alpha_M$.}
    \label{tab:alpha_M}
\end{table}

For $\alpha_M$, in the no cosmic opacity scenario, the expected precision in the first three redshift bins is around 20\%, in both realistic and optimistic cases, and grows up to ${\cal O}(1)$ for higher $z$. We summarize in Table~\ref{tab:alpha_M} some of the main current measurement of $\alpha_M$, for comparison. As compared to the CMB measurements, our proposed methodology seems to be less precise but much more accurate due to its model-independence: no particular form for the time evolution of this quantity needs to be assumed nor any considerations regarding other aspects of the gravity theory (namely, $\alpha_B$, $\alpha_K$). Compared to the measurement from GW170817, our result for the lowest bin is two orders of magnitude more precise. We note that while our forecasts are not as precise as the results from Lunar Laser, our approach allows assessing $\alpha_M$ not locally but in a wide redshift range. 

Table~\ref{tab:MB-H0} summarizes our forecasts for the Hubble parameter and the SN intrinsic magnitude $M_B$. We can achieve good precision both allowing or not for cosmic opacity. In particular, for $H_0$, assuming MG but no opacity yields a precision which is greater than that of Planck 2018 when one assumes $\Lambda$CDM~\cite{Planck:2018vyg}. For these parameter there is very little difference between the realistic and optimistic scenarios, indicating that our proposed methodology could yield precise results even with smaller datasets. We note that in traditional SN analyses, both $M_B$ and $H_0$ are completely degenerate, which means that the current $H_0$ tension could be due to a $M_B$ tension~\cite{Camarena:2021jlr}. In this approach instead one could measure both parameters independently and with good precision.

In all cases the different bins are correlated. This is due both to the correlations in $D_A$ (here forecasted to be on average 18\%), and to the fact that we use the first redshift bin to constrain $M_B$ and $H_0$. These correlations are strong for the $\eta$ bins and moderate for the $\zeta$ bins. To wit, the average correlations among $\eta$ in all bins for the case with of KN + GRB counterparts are 0.83 in both realistic and optimistic scenarios. For $\zeta$ the average is instead 0.51 (0.59) in the realistic (optimistic) scenario. In the no cosmic opacity case, the $\zeta$ correlations diminish to 0.16 (0.20) in the realistic (optimistic) scenario.

\begin{table}[t]
    \small
    \centering
    \setlength{\extrarowheight}{3pt}
    \begin{tabular}{l|c|c}
    \hhline{= = =}
    Case  & $\sigma_{H_0}$ (km/s/Mpc) realistic (optimistic) &  $\sigma_{M_B}$ (mmag) real. (opt.)  \\  		
    \hline
    1. general & 1.72 (1.71) & 0.757 (0.713)   \\
    2. no cosmic opacity & 0.47 (0.46) &  0.757 (0.713)\\
    3. GR and no opacity  & 0.35 (0.35) & 0.279 (0.239)  \\
    \hline    
    \end{tabular}
    \caption{
   Forecast uncertainty for $H_0$ and $M_B$ (68\% CI) for the GW+KN+LSS+SN dataset and different combinations of assumptions. Adding GRBs barely changes any of these numbers.}
    \label{tab:MB-H0}
\end{table}

\subsection{Forecasts for parametrized models}

We now discuss about how our model-independent results translate to constraints to parametrized models. For the cosmic opacity parameter, we consider the parametrization \cite{Avgoustidis:2009ai}
\begin{equation}
    \eta(z) = (1 + z)^{\epsilon}\,,
\end{equation}
that recovers the reciprocity-relation when $\epsilon = 0$. Using this value as fiducial model (no cosmic opacity), the forecasts of the first panel of Figure~\ref{fig:error_xi_eta} result in 
\begin{equation}
    \sigma(\epsilon) = 0.011 \quad (68\% \;\text{CI})\,.
\end{equation}
Using the parametrizations of \cite{Holanda:2012at} instead,
\begin{equation}
    \eta(z) = 1 + \eta_0z\,, \quad \text{and} \quad \eta(z) = 1 + \eta_0\frac{z}{1+z}\,,
\end{equation}
we obtain, respectively,
\begin{equation}
    \sigma(\eta_0) = 0.006\,, \quad \text{and} \quad  \sigma(\eta_0) = 0.019\,.
\end{equation}
This shows that even with high correlations between $\eta$ in different bins we could constrain to the percent level (or less) one-parameter models of cosmic opacity with this test.

For the modified gravitation wave friction, a common parametrization is \cite{Belgacem:2018lbp}:
\begin{equation}\label{eq:Xi-of-z}
    \zeta(z) = \Xi_0 + \frac{1 - \Xi_0}{(1 + z)^n}\,.
\end{equation}
The parameter $\Xi_0$ is the asymptotic value of $\zeta$ at high redshifts, while $n$ tracks how fast it goes to this asymptotic regime, which is the behavior that occurs in various modified gravity models \cite{Belgacem2019, Matos:2021qne}.

One problem with this parametrization, as noticed in \cite{Matos:2021qne}, is that General Relativity is recovered in two ways, when $\Xi_0 = 0$ or when $n=0$, a degeneracy that creates difficulties for applying this model to real data if the Universe is close to the standard model. We thus follow most of the works in the literature that consider this model and fix the slope to the value $n=2$, to track only the amplitude of the effect. Using the fiducial model with no effect ($\Xi_0$ = 1) and assuming no cosmic opacity, using the errors from GWs+KNe until $z=0.55$ and  GWs+GRBs after that, in the optimistic scenario (see Figure~\ref{fig:error_xi_no_co}), we forecast 
\begin{equation}
    \sigma(\Xi_0) = 0.009 \quad (68\% \; \text{CI})\,. 
\end{equation}

For viable $f(R)$ theories, the parameter $\Xi_0$ translates to $f_{R0}/2$, in which case the above error implies $\sigma(f_{R0}) = 0.016$, which is compatible with the result of \cite{Matos:2021qne} with fixed cosmological parameters. The above result indicates still viable modified gravity models, such as the RT non-local model with $\Xi_0 \simeq 1.8$ \cite{Belgacem:2020pdz}, could be ruled-out in the future. Finally, as far as we know, the best measurement of $\Xi_0$ with current GW data is $\Xi_0 = 1.2 \pm 0.7$ from \cite{Mancarella:2021ecn}. This means we will improve by 100 times our knowledge on this parameter with future observations, according to our analysis.

Table~\ref{tab:param_results} summarizes our results for both realistic and optimistic scenarios, and for both the general case in which both cosmic opacity and GW distance are completely free, and the case in which we assume a priori that there can be no cosmic opacity.
Our best results are obtained when considering events with KNe up to the redshifts in which they give more precise GW distances, and for higher redshifts, events with GRBs (third and sixth rows).

\begin{table}[t]
    \small
    \centering
    \setlength{\extrarowheight}{3pt}
    \begin{tabular}{c|c|c}
    \hhline{= = =}
    Data  & $\sigma_{\Xi_0}$ realistic (optimistic) & case\\  		
    \hline
    GW+KN+LSS+SN & 4.4\% (3.6\%) & general\\
    GW+GRB+LSS+SN & 5.5\% (4.0\%) & general\\
    GW+KN+GRB+LSS+SN & 3.8\% (3.2\%) & general\\
    \hline
    GW+KN+LSS+SN & 1.2\% (1.0\%) & no cosmic opacity\\
    GW+GRB+LSS+SN & 4.6\% (2.5\%) & no cosmic opacity\\
    GW+KN+GRB+LSS+SN & 1.2\% (0.9\%) & no cosmic opacity\\
    \hline
    GWTC-3+BBH mass feature & 70\% (current data \cite{Mancarella:2021ecn}) & no cosmic opacity \\
    \hline
    \end{tabular}
    \caption{Errors in the parameter $\Xi_0$ (68\% CI) from the different combinations of data/assumptions, assuming $n=2$ in Eq.~\eqref{eq:Xi-of-z}.}
    \label{tab:param_results}
\end{table}

\section{Discussion}

The goal of this work was to propose a new test which serves both as consistency-check to $\Lambda$CDM and as probe of new physics, and to investigate to which precision future GW observatories and large-scale structure surveys could perform it. This tripartite test of distances probes, with minimal assumptions, deviations from the standard relations between three cosmological distances: GW, luminosity and angular diameter. We propose a model-independent approach in which both luminosity and GW distances can exhibit deviations from their relative redshift behavior in $\Lambda$CDM. Such a difference can appear in modifications of gravity, but also when other physical phenomena take place, for instance, if the transparency of the Universe is broken -- as long as the geometric angular diameter distance is preserved.

We found that the ratio $\zeta$ of GW and luminosity distances  can be tested to a few percent in ten redshift bins up to $z = 1$, and with sub-percent error if one assumes the DDR. In addition, the cosmic opacity parameter $\eta$ will be measured to within few percent in 20 different bins up to $z=2$. The consequences for the GW friction $\alpha_M$ in scalar-tensor theories but also other modified gravity models, is that $\alpha_M > \mathcal{O}(1)$ for $z<1$ could be detected in the near future with this test. Assuming no cosmic opacity, the sensitivity to $\alpha_M$ improves by a factor of almost 2.

As usual, more precision can be obtained assuming specific parametrized models. In that case, the key parameter $\Xi_0$ can be much better constrained, reaching less than 1\% uncertainty. In fact, one could even constrain $\Xi_0$ with a combination of a bright siren Hubble diagram and other cosmological probe such as the CMB, as long as one uses a parametrization which is restrictive enough, particularly regarding the dark energy equation of state (see, e.g.~\cite{Matos:2022uew, Matos:2021qne}). The main advantage of testing distance ratios using independent distance probes is not having to rely on arbitrary choices, in particular regarding properties of dark energy or the early universe, while at the same time avoiding degeneracies with cosmological parameters. What we have shown here is that very precise results can still be obtained, even with a completely free cosmology. 

The major source of uncertainty regarding our forecasts is the fact that the amount of multi-messenger events is highly dependent on the exact electromagnetic facilities that will be operating, the amount of time they will dedicate to GW follow-up observations and the true BNS merger rate, a function that is highly uncertain from current LIGO/Virgo data~\cite{LIGOScientific:2021psn}. 

We focused on forecasts for bright standard siren observations by the future GW detector Einstein Telescope. However, other facilities planned for the next decade might also detect bright sirens, either jointly with ET or by seeing completely different events. In particular, Cosmic Explorer \cite{Reitze:2019iox} is a proposal which will likely have a sensitivity comparable to ET, and thus we expect strong overlap in the observed events and similar results for the distance errors.

The future space-based GW interferometer LISA \cite{LISA:2017pwj} is also expected to see multi-messenger events that could provide simultaneous distance and redshift measurements, and at much higher redshifts ($z \lesssim 10$). These will be most likely generated by supermassive black hole binaries (SMBHB), but forecasts for this type of events are highly dependent on several assumptions, such as the formation channel of SMBHs and the type and modelling of the electromagnetic counterparts. From the most updated forecasts of \cite{Mangiagli2022} for the event rates, it is still unclear whether there will be any multi-messenger detection at redshifts where we also see SNe or other cosmological standard candles. However, if that is the case, as we mentioned before, with bright sirens at high redshifts where we do not expect dark energy models to differ much from GR, we could replace our assumption that $D_{\rm GW} = D_{\rm L}$ for very low redshifts for the one that $M_{\ast}^2$ reaches a constant value above a certain $z = z_{\ast}$, allowing arbitrary MG models at low $z$.

Bright standard sirens can in principle also be seen with pulsar timing arrays (PTA). In fact, even if with the current observations  PTA are not capable of separating individual GW events from the stochastic background \cite{Agazie2023}, it is expected that in the future nearby SMBHBs will be identified with this technique. With such detections, not only the parameters of the binaries, including the distance, could be inferred, but also electromagnetic follow-up might be possible \cite{Charisi2021, Kelley2019}. In this case such observations could result in improved constraints on~$\zeta$.

\section*{Acknowledgements}

We thank Dr.~Josiel de Souza for useful discussions on GWDALI. ISM thanks Fundação de Amparo à Pesquisa do Estado de São Paulo (FAPESP) for the PostDoc fellowship nº 2023/02330-0. 
MQ is supported by the research agencies Fundação Carlos Chagas Filho de Amparo à Pesquisa do Estado do Rio de Janeiro (FAPERJ) project E-26/201.237/2022, Conselho Nacional de Desenvolvimento Científico e Tecnológico (CNPq) and  Coordenação de Aperfeiçoamento de Pessoal de Nível Superior - Brasil (CAPES).
RS acknowledges support from FAPESP grant n. 2021/14335-0 and 2022/06350-2, and by CNPQ grant n. 310165/2021-0 . This study was financed in part by CAPES - Finance Code 001. We acknowledge support from the CAPES-DAAD bilateral project  ``Data Analysis and Model Testing in the Era of Precision Cosmology''. MK acknowledges funding by the Swiss NSF. LA acknowledges support from DFG project  456622116. LA and MQ thank Massimo Pietroni and Marco Marinucci for collaboration on the FreePower method, and Tjark Harder for help in debugging the code.

\appendix
\section{Short review of the FreePower method}
\label{sec:freepower}

In FreePower \cite{Amendola:2023awr}, we adopt the following expression for the one-loop power spectrum \cite{Ivanov:2019pdj,DAmico:2019fhj}
\begin{align}
    P_{gg}(k,\mu,z) & =S_{{\rm g}}(k,\mu,z)^{2}\left[P^{{\rm lin}}(k,\mu,z)+P^{{\rm 1loop}}(k,\mu,z)+P^{{\rm UV}}(k,\mu,z)\right]+P^{{\rm SN}}(z)\,,\label{Pgg}
\end{align}
where $\mu$ is the cosine angle between the wavevector $\vec k$ and the line of sight.
The full expressions for $P^{{\rm 1loop}}(k,\mu,z)$
and $P^{\rm UV}$ are given in Appendix~A of \cite{Amendola:2022vte}  (see also~\cite{Ivanov:2019pdj,DAmico:2019fhj}). The linear matter power spectrum is
\begin{equation}
    P^{{\rm lin}}(k,\mu,z)=Z_1({\mathbf{k}};z)^2 G(z)^2 P_0(k)\,,\label{plin}
\end{equation}
where $P_0(k)$ is the linear  spectrum  in real space  at a fixed $z=0$,  and $G(z)$ the linear growth factor, normalized as $G(0)=1$. The RSD factor is $Z_1=b_1+f\mu^2$, where $b_{1}(z)$ is the linear bias parameter and $f(z)\equiv\dd\log G/\dd\log a$ is the linear growth rate. 

The kernels that enter the one-loop correction have been derived imposing only general symmetries, namely the equivalence principle and the generalized Galileian invariance and are therefore not restricted to Einstein-deSitter or $\Lambda$CDM cosmologies. They include five
free $z$-dependent functions, denoted $
    d_\gamma^{(2)}(z)\,,
    a_\gamma^{(2)}(z)\,,\;d_{\gamma a}^{(3)}(z)\,,
\,c_\gamma^{(2)}(z),\,a_{\gamma a}^{(3)}(z)\,.$ These functions are taken as free parameters for each $z$-bin. Moreover, the spectrum includes two bias functions, $b_1(z)$ and $b_2(z)$,  a counterterm parameter $c_0(z)$, and shot noise $P^{SN}$. An  overall smoothing factor $S_{{\rm g}}(k,\mu,z)^{2}$ takes into account both the Finger-of-God (FoG) effect and the spectroscopic errors~\cite{Euclid:2019clj,BOSS:2016psr, Amendola:2022vte}:
\begin{equation}
    S_{{\rm g}}(k,\mu,z) = \exp\left[-\frac{1}{2}(k\mu\sigma_{{\rm z}})^{2}\right] \, \exp\left[-\frac{1}{2}(k\mu\sigma_{f})^{2}\right],\label{eq:FoG}
\end{equation}
where
\begin{equation}
    \sigma_{{\rm z}}=\sigma_{0}(1+z)H(z)^{-1}\,.
\end{equation}
We take  $\sigma_{0}=0.001$ for the spectroscopic errors, and leave the FoG smoothing $\sigma_f$  as a free parameter in each bin.

The tree-level bispectrum is 
~\cite{Scoccimarro:1999ed, Desjacques:2016bnm, Yankelevich:2018uaz}
\begin{equation}
    B(\mathbf{k}_{1},\mathbf{k}_{2},\mathbf{k}_{3}) = 2\big[Z_{1}(\mathbf{k}_{1})Z_{1}(\mathbf{k}_{2})Z_{2}(\mathbf{k}_{1},\mathbf{k}_{2})G^4 P_1(k_1)P_1(k_2)+2\;{\rm cycl.}\big] 
    + B^{\rm SN} (\mathbf{k}_{1},\mathbf{k}_{2},\mathbf{k}_{3}) \label{eq:bisp}
\end{equation}
with $\mathbf{k}_{3}=-\mathbf{k}_{1}-\mathbf{k}_{2}$, and the shot noise part is
~\cite{Philcox:2021kcw,Yankelevich:2018uaz} 
\begin{equation}
    B^{\rm SN}(\mathbf{k}_{1},\mathbf{k}_{2},\mathbf{k}_{3}) = \frac{1}{n(z)} \Big[P^{{\rm lin}}(\mathbf{k}_{1})+P^{{\rm lin}}(\mathbf{k}_{2}) + P^{{\rm lin}}(\mathbf{k}_{3})\Big] \Big[1+B_{\rm sn(1)}\Big] + \frac{1}{n(z)^2}\Big[1+B_{\rm sn(2)}\Big]\,.
\end{equation}
This introduces only two new parameters, the two shot noise terms $B_{\rm sn(1)},B_{\rm sn(2)}$. Beside the bias, bootstrap, counterterms, and shot noise parameters, the spectra depend on the linear power spectrum wavebands, that we take equally spaced in $\Delta k=0.01 h/$Mpc intervals from 0.01 to 0.25 $h/$Mpc, on the growth rate $f$, and on the two geometric parameters that determine the Alcock-Paczyński (AP) effect. 

This effect introduces a  distortion of angles and distances for any cosmology that differs from the true  one, given by $\mu=\mu_{r}h /\alpha$ and  $k=\alpha k_{r}$, where ~\cite{2000ApJ...528...30M}
\begin{equation}
    \alpha\,=\,\frac{1}{l}\sqrt{\mu_{r}^{2}(h^{2}l^2-1)+1}\,.\label{eq:alpha}
\end{equation}

The distortion depends therefore on $h\equiv E(z)/E_r$ (not to be confused with the same symbol that denotes the Hubble constant today in units of 100 km/sec/Mpc) and $l\equiv L_A/L_{Ar}$, where $E(z)\equiv H(z)/H_0$ is the dimensionless Hubble function and $L_A\equiv H_0 D_A$ the dimensionless comoving angular diameter distance, and a subscript $r$ refers to the reference value (i.e. the value employed to derive the vector $k,\mu$ from the raw data). Notice that the AP effect does not measure the absolute distance $D_A$ but only the dimensionless distance $L_A$.

It is important to remark that   $l$ is actually degenerate with the combination $b_1^2 G^2(z) P_0(k)$  in the linear regime, if, as we do in the FreePower method, we want to keep the bias, growth rate, and the spectrum free to vary. However, the non-linear corrections add further dependence on $f$, $b_1$ and $P_0(k)$, and this breaks the degeneracy. Therefore, both $h$ and $l$ become measurable quantities.

\setlength\tabcolsep{3.7pt}

\begin{table}[]
    \small
    \centering
    \begin{tabular}{cccccccccccc}
    \hline
    $z$ & $V$  & $n_g\times 10^{-3}$ & $b_1$ & $b_2$ & $c_0$ & $a^{(2)}_\gamma$ & $c^{(2)}_\gamma$  & $d^{(2)}_\gamma$ & $a^{(3)}_{\gamma a}$ & $d^{(3)}_{\gamma a}$ &$\sigma_f$ \\
    &\!\![Gpc$/h]^3\!\!$&  $[h/{\rm Mpc}]^{3}$ & & & $\!\![{\rm Mpc}/h]^2\!$ &&&&&& $\![{\rm Mpc}/h]\!$\\
    \hline
 0.05 & 0.0356 & 236 & 1.38 & -2.14 & -22.9 & 1.43 & 2.18 & 0.872 & 0.665 & 0.286 & 3.34 \\
 0.15 & 0.228 & 50.2 & 1.45 & -2.14 & -22.9 & 1.43 & 2.33 & 0.87 & 0.665 & 0.286 & 3.44 \\
 0.25 & 0.557 & 23.8 & 1.53 & -2.14 & -22.9 & 1.43 & 2.49 & 0.868 & 0.665 & 0.286 & 3.5 \\
 0.35 & 0.972 & 8.11 & 1.61 & -2.14 & -22.9 & 1.43 & 2.65 & 0.866 & 0.665 & 0.286 & 3.52 \\
 0.45 & 1.43 & 2.27 & 1.69 & -2.14 & -22.9 & 1.43 & 2.82 & 0.865 & 0.665 & 0.286 & 3.5 \\
 0.55 & 1.90 & 0.411 & 1.78 & -2.14 & -22.9 & 1.43 & 2.99 & 0.864 & 0.665 & 0.286 & 3.47 \\
 0.65 & 2.86 & 1.51 & 1.1 & -0.798 & -53 & 1.43 & 1.63 & 0.858 & 0.665 & 0.286 & 3.4 \\
 0.75 & 3.33 & 2.37 & 1.2 & -0.837 & -53 & 1.43 & 1.83 & 0.858 & 0.665 & 0.286 & 3.4 \\
 0.85 & 3.76 & 1.77 & 1.24 & -0.847 & -53 & 1.43 & 1.91 & 0.858 & 0.665 & 0.286 & 3.27 \\
 0.95 & 4.14 & 1.34 & 1.28 & -0.852 & -53 & 1.43 & 1.99 & 0.858 & 0.665 & 0.286 & 3.27 \\
 1.05 & 4.48 & 1.01 & 1.32 & -0.852 & -53 & 1.43 & 2.07 & 0.858 & 0.665 & 0.286 & 3.1 \\
 1.15 & 4.77 & 0.777 & 1.36 & -0.847 & -53 & 1.43 & 2.15 & 0.858 & 0.665 & 0.286 & 3.1 \\
 1.25 & 5.01 & 0.598 & 1.4 & -0.843 & -53 & 1.43 & 2.23 & 0.858 & 0.665 & 0.286 & 2.93 \\
 1.35 & 5.22 & 0.449 & 1.44 & -0.838 & -53 & 1.43 & 2.31 & 0.858 & 0.665 & 0.286 & 2.93 \\
 1.45 & 5.4 & 0.319 & 1.48 & -0.828 & -53 & 1.43 & 2.39 & 0.858 & 0.665 & 0.286 & 2.77 \\
 1.55 & 5.56 & 0.23 & 1.52 & -0.813 & -53 & 1.43 & 2.47 & 0.858 & 0.665 & 0.286 & 2.77 \\
 1.65 & 5.74 & 0.175 & 1.56 & -0.798 & -53 & 1.43 & 2.55 & 0.858 & 0.665 & 0.286 & 2.62 \\
 1.75 & 5.96 & 0.13 & 1.6 & -0.779 & -53 & 1.43 & 2.63 & 0.858 & 0.665 & 0.286 & 2.62 \\
 1.85 & 6.26 & 0.0998 & 1.64 & -0.754 & -53 & 1.43 & 2.71 & 0.858 & 0.665 & 0.286 & 2.47 \\
    \hline
    \end{tabular}
    \caption{ Our forecast specifications and fiducials for the joint DESI-like + Euclid-like survey. }
    \label{tab:DESI-Euclid}
\end{table}

The fiducial for the growth rate is taken to be $f=\Omega_m(z)^{0.545}$, and for $P(k),H,D_A$ we adopt $\Lambda$CDM values: $\Omega_{c}=0.270$,
$\Omega_{b}=0.049$, $\Omega_{k}=0$, $h=0.67$, $n_{s}=0.96$, and
$\sigma_{8}=0.83$. The fiducials for all the other parameters are listed in Table \ref{tab:DESI-Euclid}. We assume infinite prior (which, in the Fisher formalism, means no prior) for all parameters, except a 100\% prior on four parameters: $\sigma_f$, $1+P_{\rm sn}$ and $1+B_{\rm sn(1,2)}$.  We do not include the covariance between spectrum and bispectrum because it has been estimated to give a negligible contribution~\cite{Yankelevich:2018uaz,Amendola:2023awr}. The  Fisher matrix  $\mathbf{F}$ is given by 
\begin{equation}
    F_{\alpha\beta}\,=\, \sum_{i,j} X_{i,\alpha}C_{ij}^{-1}X_{j,\beta}
\end{equation}
where the data vector $X_{i}=\{P,B\}$ includes the spectrum $P$
for every  $k,\mu$ bin, and the bispectrum $B$ for every $k_1,k_2,k_3,\mu_1,\mu_2$ bin, and $()_{,\alpha}$ represents the partial derivative with respect to parameter $\alpha$ calculated on the fiducial. The correlation matrix $C_{ij}$ is given by two blocks. The $P$ block is 
\begin{equation}
    C_{PP,ij}=2P_{gg}(\mathbf{k}_{i})P_{gg}(\mathbf{k}_{j})\delta_{ij}/N_{P}\,,
\end{equation}
where
\begin{equation}
    N_{P}=\frac{V}{(2\pi)^{2}}k_{i}^{2}\Delta k\Delta\mu
\end{equation}
gives the number of $\vec k$ vectors in a $k,\mu$ bin. The $B$ block is  (see e.g.~\cite{Fry:1992ki,Scoccimarro:2003wn,Yankelevich:2018uaz})
\begin{equation}
    C_{BB,ij}=s_{B}\frac{V}{N_{B}}G^6 P_1(k_{1})P_1(k_{2})P_1(k_{3})\delta_{ij}l_{i}\label{eq:cbb}\,,
\end{equation}
where $s_{B}=6,2,1$ for equilateral, isosceles, and scalene triangles, respectively, and  $l_{i}=2$ for co-linear triangles and 1 otherwise \cite{Chan:2016ehg}. The number of triangular configurations per bin is 
\begin{equation}
    N_{B}=2\frac{V^{2}}{8\pi^{4}}k_{i_1}k_{i_2}k_{i_3}(\Delta k)^{3}\Sigma(\Omega)\Delta\Omega\,,
\end{equation}
where $k_{i_{1,2,3}}$ are the central value of the $k$-bins, and $\Sigma(\Omega)\Delta\Omega$ is the number of triangles within the angle orientation $\Delta\Omega=(\Delta\mu)^{2}$,
that depends on which coordinate system is used.  For both $P$ and $B$, we adopt angular bins of size $\Delta\mu=0.1$.

\bibliographystyle{JHEP}
\bibliography{sample,references}

\end{document}